\documentclass[onecolumn,showpacs,preprintnumbers,amsmath,amssymb]{revtex4}
 \usepackage{graphicx}
\usepackage{latexsym}  
\usepackage{dcolumn}
\usepackage{graphicx}

\usepackage{graphicx}% Include figure files
\usepackage{dcolumn}% Align table columns on decimal point
\usepackage{bm}% bold math

\usepackage{bm}

\begin{document}

\newcommand{\eq}{\begin{equation}}                                                                         
\newcommand{\eqe}{\end{equation}}             

\title{Self-similar shock wave solutions of the non-linear Maxwell equations}

\author{ I. F. Barna}
\address{Wigner Research Center of the Hungarian Academy 
of Sciences, \\  H-1525 Budapest, P.O. Box 49, Hungary \\ 
and \\ 
ELI-HU Nonprofit KFT, H-6720 Szeged, Dugonics t\'er 13., Hungary}
% $^b$University of P\'ecs, PMMK, Department of Mathematics and Informatics, 
% Boszork\'any u. 2, P\'ecs, Hungary}
1
\date{\today}

%%%%%%%%%%%%%%%%%%%%%%%%%%%%%%%%%%%%%%%%%%%%%%%%%%%%%%%%%%%%%%%%%%%%%%%
\begin{abstract} 
       In our study we consider non-linear, power-law field-dependent electrical permittivity and magnetic permeability and investigate the  time-dependent Maxwell equations with the self-similar Ansatz. This is a first-order hyperbolic PDE system which can conserve non-continuous initial conditions describing electromagnetic shock. Such phenomena may happen in complex materials induced by the planned powerful Extreme Light Infrastructure(ELI) laser pulses. 
 
\end{abstract}

%\draft
\pacs{03.50.De, 02.30.Jr}
% Other topics in heat transfer, partial differential equations   
\maketitle
%%%%%%%%%%%%%%%%%%%%%%%%%%%%%%%%%%%%%%%%%%%%%%%%%%%%%%%%%%%%%%%%%%%%%%%                                                  
\section{Introduction}
Wave propagation in non-linear media is a fascinating field in physics with large literature \cite{with}. To study such effects diverse non-linear partial differential equations(PDEs) have to be investigated with various methods.
One of the best known non-linear wave propagation phenomena is the solitary wave, usually based on the non-linear 
Schr\"odinger or sine-Gordon or KdV equations. On the other side there are many more, not-so well-known non-linear wave equations  exist which have delicate properties such as, shock-waves, solutions with continuous compact support and so on. 
Such equations are the various Euler or unconventional heat conduction equations.     
 To investigate if a system has such properties, one of the most powerful analytical tool is to apply the self-similar Ansatz which may describe dispersive solutions with reasonable physical interpretation. The validity of such solutions is very wide in continuum mechanics and mostly used to study shock-waves and other fluid dynamical problems \cite{sedov, zeld,barenb}.

           In one of our former studies we investigated the paradox of heat conduction with a new kind of time-dependent Cattaneo heat conduction law \cite{barn} and found physically reasonable solutions with compact support. In another analysis we presented three dimensional analytical results for the Navier-Stokes equations \cite{barn2}. 
The properties of the self-similar solution will be analyzed later. 
    
From the four Maxwell field equations combining with the two 
constitutive relations a linear second-order hyperbolical wave equation can be derived for the field variables. In such cases the constitutive equations contain only linear relations for the electrical permittivity and for the magnetic permeability. 
The theory of the electromagnetic wave propagation can be found in various textbooks \cite{papas}. 

Under non-linear Maxwell equation most of the people mean the non-paraxial non-linear Schr\"odinger equation (NNSE) 
which is derived from the Helmholtz equation including the Kerr media where the relative dielectric permittivity is well 
described $\epsilon_r  = n^2 = n_0^2 + \delta_{NL}(E_y)$. Here $n_0$ is the linear contribution to the total refractive index n. In sufficiently slow media, where the characteristic response time of the non-linearity is much greater than the temporal period of the field oscillations one has $\delta_{NL} \approx 2n_0^2n_2 <E_y^2>$ where $n_2$ is the Kerr coefficient 
and $<>$ denotes the time averaging over many optical cycles.   

There are large number of studies available where the NNSE is analytically (or numerically) solved and analyzed. Additional literature can be found in \cite{sanch}. 

When ultra short intense laser pulses propagate in a media
then there is an intensity dependence of the group velocity which leads to 
the phenomena of self-steepening and optical shock wave formation.  
It means that the peak of the pulse is slowed down more than the 
edge of the pulse, leading to steepening of the trailing edge of the pulse. 
The envelope becomes steeper and steeper. 
If the edge becomes infinitely steep, it is said to form an optical shock wave. 
Self-steepening has been described by various authors \cite{boyd, de, yang,roth,ranka,gudz}. 
In non-linear media optical beams can suffer self-trapping where the wave equation is solved with the displacement field of 
$D = \epsilon_0 E + \eta E^3$ \cite{chiao,rosen}. 

Most of the authors consider electromagnetic shock waves in this sense.

We however follow a different way, in our recent study we consider non-linear, power-law field-dependent electrical permittivity and magnetic permeability and investigate the last two time-dependent Maxwell equations with the self-similar Ansatz. This is now a first-order hyperbolical non-linear PDE system which can conserve non-continuous initial conditions describing electromagnetic shock-waves.
To our knowledge the question of electromagnetic shock-wave is only laconically shortly mentioned in a well-known physic textbook \cite{landau}. Onyl a small number of publications exist (most of them are from Russian authors) about electromagnetic shock wave propagations is anisotropic magnetic materials where the direct Maxwell equations  \cite{gvoz}  are solved. 
Shock waves in transition lines are investigated in the book of \cite{sch}.
 Unfortunately, not with our method. 

Direct integration of the Maxwell equations for dielectric resonators is a new research context 
for future novel particle accelerators \cite{schess}.  Such effects might happen in complex materials which could be induced by powerful laser pulses which will be available in the planned  Extreme Light Infrastructure (ELI).  

%%%%%%%%%%%%%%%%%%%%%%%%%%%%%%%%%%%%%%%%%%%%%5
\section{Theory and Results} 
Let's start with  the usual four Maxwell equations for the fields:

\begin{eqnarray}
 \nabla \cdot {\bf{D}} = \rho,   \hspace*{1cm}  \nabla \cdot {\bf{B}} = 0,  \nonumber \\  
\nabla  \times {\bf{E}}  =- \frac{\partial {\bf{B}}} {\partial t }, \hspace*{1cm} \nabla  \times {\bf{H}} 
=\frac{\partial {\bf{D}}} {\partial t}  + {\bf{J}}, 
\label{maxw} 
\end{eqnarray}
where ${\bf{E,B}}$ are the electric and magnetic fields ${\bf{D,H}}$ are the electric displacement and magnetizing fields 
$\rho$ is the electric charge and ${\bf{J}}$ is the current density which is zero in insulator media. 
The closing constitutive relations are  
\eq 
{\bf{D}} = \epsilon {\bf{E}},  \hspace*{1cm}  {\bf{B}} = \mu {\bf{H}},   
\label{const}
\eqe
where $\epsilon$ is the electrical permittivity and $\mu $ is the magnetic permeability. 
For non-isotropic linear materials $\epsilon$ and $\mu$ are second order tensors. 
($D_{\alpha} = \sum_{\beta} \epsilon_{\alpha\beta}E_{\beta}$ and $ H_{\alpha} = \sum_{\beta}\mu_{\alpha\beta}B_{\beta}$).   
For the linear and isotropic materials these are pure real numbers. 
The most general linear relation for the constitutive equations is however the following
\eq
D_{\alpha}({\bf{x}},t) = \sum_{\alpha}\int d^3x' \int dt' {\bf{\epsilon}}_{\alpha,\beta}({\bf{x'}},t')  
 E_{\beta}({\bf{x'-x}},t-t').
\eqe
(In Eq. 3 the  $D_{\alpha} \rightarrow B_{\alpha} $ and 
$ E_{\beta} \rightarrow H_{\beta} $ interchange is still valid.) 
This equation means non-locality both in space and time. The later can be addressed as memory effects, too.
The Fourier transform of the electrical permittivity is the frequency dependent dielectric function, which attracts 
much interest. The crucial symmetry properties can be expressed via the Kramers-Kronig formula \cite{landau,jacks}
which defines the relation between the real and the imaginary part. 

 Equations of (\ref{maxw}-\ref{const}) are enough to derive the usual second-order liner hyperbolic wave-equation for 
the field variables, which can be found in any electrodynamics textbook \cite{jacks}. 
However, this classical calculation is based on a numerical trick, an additional spatial derivation is done, which also means  that the first derivatives of the 
fields are continuous and small.  But the original Maxwell equations are of first order both in 
time and space, therefore the initial conditions does not need to be continuous.  This is a crucial point and the main motivation of our analysis. 

According to the basic book of Zel'dovich and Raizer \cite{zeld} to describe the propagation of  {\it{large}} mechanical disturbances (non-continuous tears, shock waves) in a media the first order hyperbolic Euler and continuity equations have to be applied. These equation also have $f(x-\tilde{c}t)$ traveling wave solutions with a velocity of $\tilde{c}$ which is larger then the propagation of sound.  The speed of sound however enters the gas dynamic equations. 
 On the other side the propagation velocity of {\it{small}} mechanical disturbances can be described via second-order wave equations.
 In this language we may speak about two different kind of wave equations or wave propagation phenomena.  
The Maxwell equations should be considered for large electromagnetic disturbances and the second order wave equation for the small  (e.g. sinusoidal) electromagnetic disturbances.  
We follow this analogy and apply non-linear material laws and solve directly the first-order hyperbolic Maxwell equations for propagation. 
 
Maxwell equations in vacuum are linear in the fields of ${\bf{ B}}$ and ${\bf{E}}$. Many hundred 
telephone conversation can propagate parallel on a single microwave link without any distortion. Another experimental 
evidence of the linearity is the idea of linear superposition. In optics white light is refracted by a prism into the color of the 
rainbow and recombined into white light again. 
There are of course, circumstances when non-linear effects occur in magnetic materials or in crystals responding to intense 
laser beams like frequency doubling. 

Our non-linear Maxwell equation is, however defined in a complete different way, namely through the following non-linear 
material ( or constitutive) equations
\eq 
 \mu({\bf{H}}) = a {\bf{H}}^{q},   \hspace*{1cm}   \epsilon({\bf{E}}) = b{\bf{E}}^{r}, 
\label{anyagegy}
\eqe

where all the four free parameters   (a,b,q,r) are real numbers (for physical reasons $\epsilon({\bf{E}}) \cdot   \mu({\bf{H}})  > 0$ ) and  the parameters 
$a$ and $b$ are present to fix the proper physical dimensions. 
(Such power law dependence of material constants are popular in different flow problems like  in heat propagation \cite{zeld} where the heat conduction constant can have temperature dependence  like $\kappa \sim T^{\nu}$.)
Note, that through these relations we define space and  time dependent(dynamical) material equations which are still local in space and time.  (We neglect now the metamaterials where 
$\epsilon$ and $\mu$ could have negative values \cite{ves}.) 
  
We know from special relativity that the speed of light in vacuum is the largest 
available wave propagation which can carry physical information, and can be evaluated from 
the electric and magnetic properties as well $ c^2 = 1/(\mu_0\epsilon_0)$. 
(The zero subscript stands for vacuum.)  
Permeability and permittivity are not fully independent from each other.  
This formula is slightly modified for any additional media like 
$ c^2_m = 1/(\mu_0 \mu_m \epsilon_0  \epsilon_m) $   where the subscript m stands for the media. 
It is also clear that any  stable electromagnetic wave propagation speed in media is always less than the speed of light 
in vacuum $ (c_m < c  )$  but for a short time quick particles (usually charged) can propagate quicker than the local speed of light producing Cherenkov radiation. 
Therefore in our calculations we will use the following relation: 
 \eq 
\mu(H) = aH^{q} , \hspace*{1cm}  \epsilon(H) = \frac{1}{c^2 a H^q}.
\label{anyagegy}
\eqe
Note, that now the propagation speed of the electromagnetic signal has an upper bound which is c.  
(From now on we will consider one spatial coordinate and neglect the vectorial notation.)
With this constrain we reduced the number of the four independent parameters to two.  
In electromagnetic wave propagation the role of $\epsilon$ and $\mu$ are symmetric, 
however we use this relation because of the existence of  {\bf{J}} in the last Maxwell equation. 
We will see later on that with this choice the ordinary differential equation which is obtained from the third 
Maxwell equation can be integrated and the solutions became more transparent. 

For the current density we apply the differential Ohm's law
\eq
{\bf{J}} = \sigma {\bf{E}},
\label{ohm}
\eqe
where $\sigma$ is the conductance of the media - and can be a second rank tensor in crystals or a highly non-linear field dependent quality like the permeability or the susceptibility  $\sigma = h{\bf{E}}^{p}.$  
In a transition-metal oxide it can be a $ \sigma  \approx (1/E)sinh(E) $ function \cite{freud}. 
 
(Our physical intuition says that only some integers  ($\mp 1, \mp 2$) and  some rational numbers  ($\mp 1/2, \mp 2/3$) 
will be crucially interesting.) 

%%%%%%%%%%%%%%%%%%%%%%%%%%%%%%%%%%%%%%%%%%%%%%%%%%%%
%\begin{figure}
%* \vspace*{1.0cm} 
%\hspace*{0.5cm}
%\scalebox{0.5}{
%\rotatebox{0}{\includegraphics{selfsim.ps}}}  
%\vspace*{-1.0cm} 
%\caption{A self-similar type of solutions Eq. (\ref{self})
% are the presented Gaussian curves for regular heat conduction.  $t_1 < t_2$ }  	
%\label{negyes}       % Give a unique label 
%\end{figure}
%%%%%%%%%%%%%%%%%%%%%%%%%%%%%%%%%%%%%%%%%%%%%%%%%%%%%% 

For the sake of simplicity we consider the following one dimensional wave propagation problem 
\eq
{\bf{E}} = (0, E_y(x,t),0), \hspace*{1cm}  {\bf{H}} = (0,0, H_z(x,t)),   
\label{koord}
\eqe
which means linearly polarized electric filed in y direction with x coordinate dependence 
and linearly polarized magnetic field in z direction with x coordinate dependence only. 
Now the last two Maxwell equations are  

\eq
\frac{\partial E_y}{\partial x} =  - \frac{\partial B_z}{\partial t},  \hspace*{1cm} 
-\frac{\partial H_z}{\partial x} =   \frac{\partial D_y}{\partial t} + J_y. 
\eqe

From basic textbooks \cite{sedov,zeld,barenb}  the form of the one-dimensional self-similar Ansatz can be taken 
\eq 
T(x,t)=t^{-\alpha}f\left(\frac{x}{t^\beta}\right):=t^{-\alpha}f(\eta) 
\label{self}
\eqe 
where $T(x,t)$ can be an arbitrary variable of a
 partial differential equation(PDE) and $t$ means time and $x$ means spatial 
dependence.
The similarity exponents $\alpha$ and $\beta$ are of primary physical importance since $\alpha$  represents the rate of decay of the magnitude $T(x,t)$, while $\beta$  is the rate of spread  
(or contraction if  $\beta<0$ ) of the space distribution as time goes on. 
 Solutions with integer exponents are called self-similar solutions of the first kind
(and sometimes can be obtained from dimensional analysis of the problem). 
The above given Ansatz can be generalized considering real and continuous functions a(t) and b(t) instead of $t^{\alpha} $ and $t^{\beta}$.   

The most powerful result of this Ansatz is the fundamental or 
Gaussian solution of the Fourier heat conduction equation (or for Fick's
diffusion equation) with $\alpha =\beta = 1/2$. 
This transformation
is based on the assumption that a self-similar solution
exists, i.e., every physical parameter preserves its
shape during the expansion. Self-similar solutions usually
describe the asymptotic behaviour of an unbounded or a far-field
problem; the time t and the space coordinate x appear
only in the combination of  $f(x/t^{\beta})$. It means that the existence
of self-similar variables implies the lack of characteristic
length and time scales. These solutions are usually not unique and
do not take into account the initial stage of the physical expansion process.
These kind of solutions  describe the intermediate asymptotics of a problem: they hold when the precise initial 
conditions are no longer important, but before the system has reached its final steady state. 
For some systems it can be shown that the self-similar solution fulfills the source type (Dirac delta) 
initial condition, but not in our next case. 
They are much simpler than 
the full solutions and so easier to understand and study in different regions of parameter space. A final reason for studying 
them is that they are solutions of a system of ordinary differential equations and hence do not suffer the extra inherent numerical 
problems of the full partial differential equations. In some cases self-similar solutions helps to understand diffusion-like properties 
or the existence of compact supports of the solution. 

Applicability of this Ansatz is quite wide and comes up in various 
transport systems \cite{sedov,zeld,barenb,barn,barn2}.  

For our problem we consider the following Ans\"atze: 
\eq
E_y(x,t)=t^{-\alpha}f\left(\frac{x}{t^\beta}\right):=t^{-\alpha}f(\eta); \hspace*{5mm} 
H_z(x,t)=t^{-\delta}g\left(\frac{x}{t^\beta}\right):=t^{-\delta}g(\eta).
\label{ans}
\eqe
Where $\alpha, \beta, \delta$  are three real number which are (at this point of the model)  independent from each other.  
The functions $f(\eta)$ and $g(\eta)$ are the shape functions of the problem.

Combining  Eq. (\ref{ans}) together with  Eq. (\ref{anyagegy}),   (\ref{ohm}) and inserting into the original last  two Maxwell equations we get the following system: 
\begin{eqnarray}
\frac{\partial}{\partial x}[t^{-\alpha}f] & = &-\frac{\partial}{\partial t}[
a t^{-\delta (q+1)}g^{q+1}   ] \nonumber \\ 
-\frac{\partial}{\partial x}[t^{-\delta}g] &  = &\frac{\partial}{\partial t}[c^{-2}a^{-1} t^{\delta q -\alpha}
 g^{-q}f]  +   ht^{-\alpha( p+1)}f^{p+1}.  
\end{eqnarray}
Having done the derivations  we arrive at the next ordinary differential equation(ODE) system 
\begin{eqnarray}
f'   & =&     a(q+1)[\delta g^{q+1} + g^qg' \eta \beta ],                 \nonumber \\ 
-g'  &=&  \frac{1}{a c^2}[(q+1) g^q f + q(q+1) g^{q-1} g' f\eta  +  (q+1)g^q f' \eta] 
\label{odesys}
\end{eqnarray}
where prime means derivation with respect to $\eta$.  
Note, that if  the following universality relations are held ($\delta =1$ and $\beta = q+1$) the first equation is a total derivative and can be integrated 
resulting 
\eq
f = a(q+1)\eta g^{q+1}. 
\label{feltetel}
\eqe
This fixes the connection between the electric and magnetic fields.   
From the second ODE we get that $\alpha = 1$ and $p=1$  should be. 
This means that our media should not have any conductivity for self-similar solutions. 
Inserting Eq. (\ref{feltetel}) into the second equation of (\ref{odesys}) 
we arrive to our final expression 
\eq
-g' = \frac{2(q+1)^2\eta g^{2q+1}  +h}{1 + (2q+1)(q+1)^2\eta^2 g^{2q}}
\label{vegso}
\eqe  
where light velocity c is fixed to unity remaining q and h the final two free parameters. We set $h$ to 0. 
 Note, that now different real q values mean different 
exponents for magnetic permeability representing different physical material properties and different physics. 

For general q only an implicit solution can be given 
\eq
g + g^{2q+1}\eta^2 q^2 +  2g^{2q+1}\eta^2 q + g^{2q+1}\eta^2  - c = 0. 
\label{sol}
\eqe
 For some exponents explicit solutions can be obtained. In general we can investigate the direction field of the ODE 
which gives us qualitative and global information of the solutions. Note, that (14) is non-autonomous (depending on $\eta$) 
therefore there is no general theorem to study the  direction field.   
A careful analysis for definite $q$ values clearly shows that there are two distinct classes of solutions available.  \\ 
For $q< -1/2$ there are some solutions with compact supports, otherwise all the solutions are continuous on the whole plain.  
For $q = -1$ there is an exception the Eq. (14) becomes trivial  and $g(\eta) = const. $    
It is clear from (14) that for $q<-1/2$ the denominator can be zero, therefore a singularity can appear where the first 
derivative of $g(\eta)$ becomes infinite. This dictates a vertical direction field. If a solution with an initial condition meets 
this field line than it stops, and cannot be continued.  These point can be calculated from the denominator. 
On figure 1a we present the direction field for $q = -2$. 
The shock front (or the compact support)  is formed on the $g(\eta) = 3^{1/4} \sqrt{\eta}$ which is easy to identify.  
The compact support of the ODE solution of Eq. (\ref{vegso}) means that the solution of the original PDE system for $E_y(x,t)$ 
is also compact via the $ \eta = x / t^{\beta}$ in real time and space.   
The constrain  (Eq. (\ref{feltetel}))  between f and g also dictates the same compact support for the $H_z(x,t)$ field as well.    
Outside these time and x coordinate ranges we may fix the values of $E_y(x,t) $ and $H_z(x,t)$ identically to zero 
which are also solutions of the last two Maxwell equations.  
In this way we can construct the shock-wave solutions for the original PDEs. 

Figure 1b shows the solution for $q=1/2$.  
For $q \ge -1/2$ the denominator cannot be zero therefore no infinite derivatives exist.   
Luckily, the explicit solution can be given $ g(\eta) =  \frac{-2c  \mp  \sqrt{4c + 9 \eta^2}}{9c\eta^2} $ which is continuous for every $\eta$.    \\ 

To find physically reasonable solutions we calculate the Poynting vector which gives us the energy flux 
(in $W/m^2$) of an electromagnetic field.  Unfortunately, there are two controversial form of the Poynting 
vector in material  based on the Abraham or the Minkowski formalism, a detailed description can be found in \cite{poy}. 
 We use the next form of the Poynting vector 
\eq
 {\bf{S}} = {\bf{E}} \times {\bf{H}} = t^{-\alpha-\delta}fg =   t^{-\alpha-\delta} a(q+1)\eta g^{q+2}. 
\eqe
Note that for $q<-2$ the  $\int_0^{cut} S d\eta$ is finite which is a good result. 
The spacial  integral of the pointing vector $ \sim \int_0^{cut} \frac{x}{t^{\beta}}g^{q+2}(\frac{x}{t^{\beta}}) dx$ 
is finite for all time. However the time integral of   $ \sim \int_0^{cut} \frac{x}{t^{\beta}}g^{q+2}(\frac{x}{t^{\beta}}) dt$ 
can be problematic for small t, and depends on the concrete form of g. 

Another method to classify if the solutions are physical would be the total energy of the fields in a finite volume.  
For linear electrodynamics the energy density is defined as follows 
$ W =  \frac{1}{2}({\bf{E}}\cdot{\bf{D}} + {\bf{B}}\cdot{\bf{H}} )
\label{energia} $  
However, even this formula is  problematic. There are several non-linear electrodynamic theories
like by Born \cite{born}  or  by Rafelski \cite{grei} 
based on the Lagrangian density where W contains additional terms. 
Kotel'nikov \cite{kot} generalized the Born model and suggested an infinite series of Lorentz and Poincar\'e-invariant non-linear versions of the Maxwell equations. 

Our approximation to describe the permeability and permittivity Eq. (\ref{anyagegy})  is just one way to a non-linear model. 
Another physically tenable description for the constitutive equations could be a series expansion like 
$  \epsilon(E) = 1+aE +bE^2 ...)  $ where the linear term is responsible for the so-called Pockels or electro-optical effect and the quadratic term is for the Kerr effect(a and b are constants to fix the proper dimension).   
For optically important materials $ \mu = const $ is the right choice.  Non-linear magnetic properties play a significant role 
only for plasmas where additional hydrodynamical equations have to be taken into account. 
Unfortunately, our well-established Ansatz  Eq. (\ref{ans})  does not applicable directly to such power series.

%%%%%%%%%%%%%%%%%%%%%%%%%%%%%%%%%%%%%%%%%%%%
\begin{figure}
    \label{fig:subfigures}
    \begin{center}
{{\scalebox{0.45}{\rotatebox{0}{\includegraphics{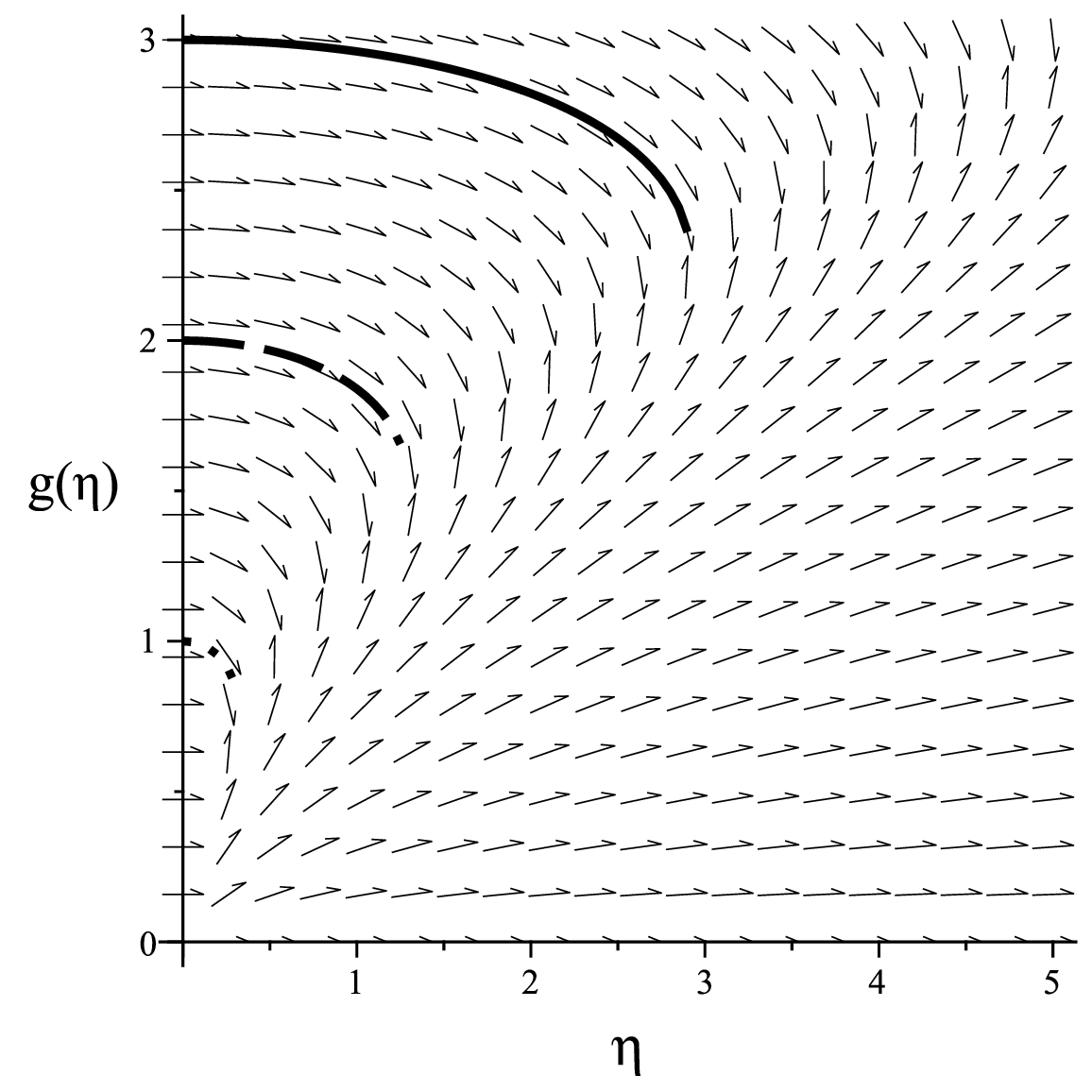}}}} }
{{\scalebox{0.45}{\rotatebox{0}{\includegraphics{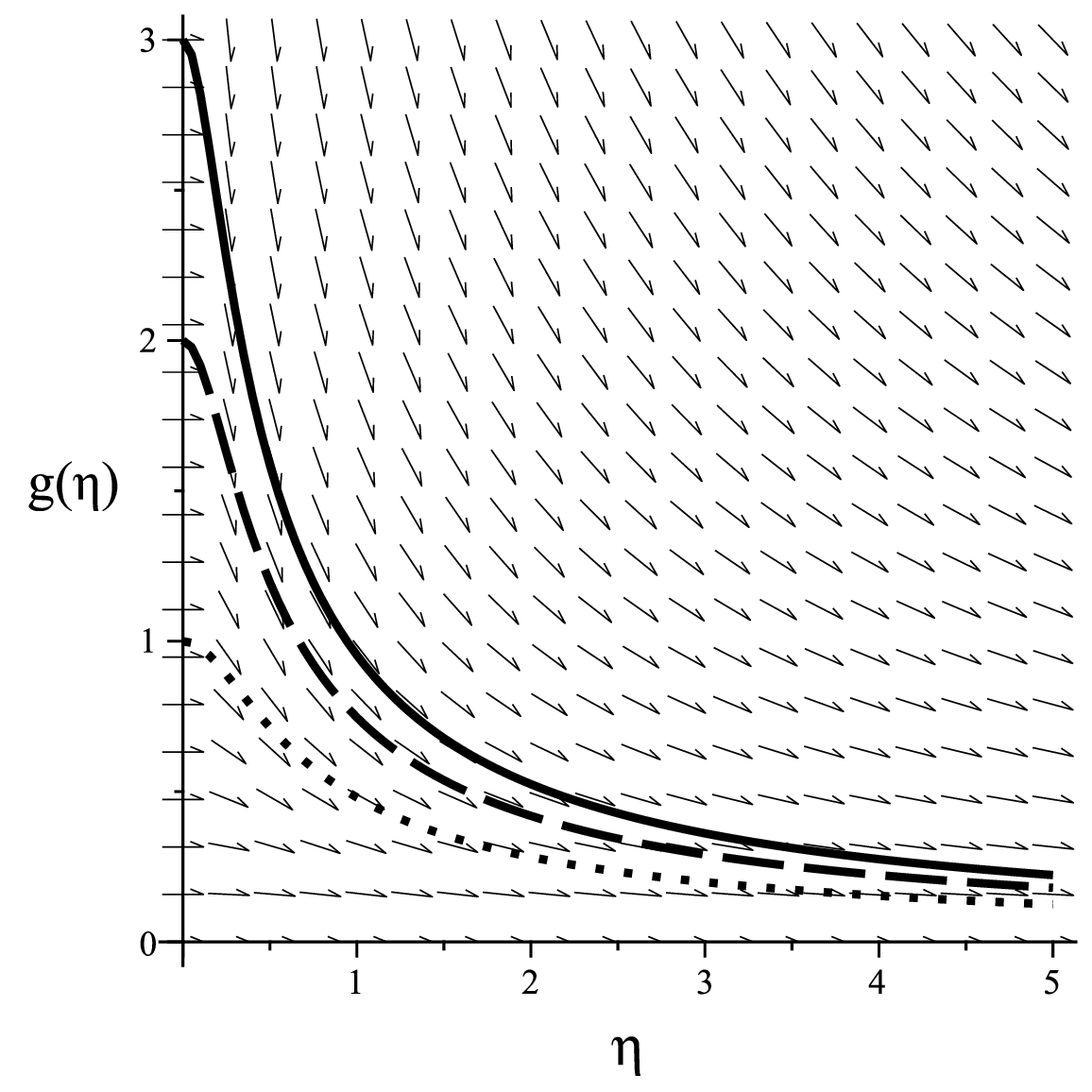}}}} }
a)   \hspace*{3cm} b)
    \end{center}
    \caption{The direction field of   Eq. (\ref{vegso})  a)    for  $q = -2$ and b)  for $q=1/2$.
 The solid, dashed and dotted lines present numerical solutions  $ f(0)=3$,   $f(0)=2$ and $f(0)=1$  initial conditions for both $q$s.
     }%
\end{figure}
%%%%%%%%%%%%%%%%%%%%%%%%%%%%%%%%%%%%%%%%%%%%

\section{Summary}
We introduced a power law magnetic field dependent magnetic permeability and investigated the 
corresponding non-linear Maxwell filed equations with the self-similar Ansatz.  
 If the power law exponent  is smaller than minus half then compact,  
 shock-wave like solutions are obtained which might have some importance in laser matter interactions. 
The work was supported by the Hungarian HELIOS project and by the Hungarian OKTA NK 101438 Grant. The paper is dedicated to my two year old daughter Annabella. 
%%%%%%%%%%%%%%%%%%%%%%%%%%%%%%%%%%%%%%%%%%%%%%%%%%%%%%%%%%%%%%%%%%%%%%                                    
                                                                  
%%%%%%%%%%%%%%%%%%%%%%%%%%%%%%%%%%%%%%%%%%%%%%%%%%%%%%%%%%%%%%%%%%%%%%%        
\end{document}